\documentclass[conference]{IEEEtran}
\IEEEoverridecommandlockouts
\usepackage{xurl}
\usepackage[hidelinks]{hyperref}

\usepackage{xurl}
\usepackage[hidelinks]{hyperref}

\usepackage{colortbl}
\usepackage{cite}
\usepackage{amsmath,amssymb,amsfonts}
\usepackage{algorithmic}
\usepackage{graphicx}

\usepackage{subcaption}
\usepackage{multirow}
\usepackage{booktabs}
\usepackage{color,soul}
\usepackage{stix}
\usepackage[table]{xcolor}

\usepackage{tcolorbox}
\usepackage{dblfloatfix}
\usepackage{booktabs}

\newtcolorbox{boxH}{
    colback = white!90!gray, 
    colframe = black, 
    boxrule = 0pt, 
    leftrule = 3pt
}

\def\BibTeX{{\rm B\kern-.05em{\sc i\kern-.025em b}\kern-.08em
    T\kern-.1667em\lower.7ex\hbox{E}\kern-.125emX}}
\begin{document}

\title{What's Inside a GitHub Repository? An Empirical Study on the Contents of 10K Projects}


 \author{
 \IEEEauthorblockN{Andre Hora}
 \IEEEauthorblockA{
\textit{DCC, UFMG}\\
 Belo Horizonte, Brazil \\
 andrehora@dcc.ufmg.br}
 \and
  \IEEEauthorblockN{João Eduardo Montandon}
 \IEEEauthorblockA{
\textit{DCC, UFMG}\\
 Belo Horizonte, Brazil \\
joao@dcc.ufmg.br}
 \and
  \IEEEauthorblockN{Diego Elias Costa}
 \IEEEauthorblockA{
\textit{REALISE Lab, Concordia University}\\
 Montreal, Canada \\
diego.costa@concordia.ca}
 }

\maketitle

\begin{abstract}
    GitHub is the largest code hosting platform, with millions of repositories spanning multiple technologies.
    Despite this, little is known about the actual contents of GitHub's repositories in the wild.
    This paper presents an initial empirical analysis to better understand the contents of real-world GitHub repositories.
    We analyze the files, directories, and extensions present in 10,000 GitHub repositories, as well as their evolution over ten years.
    Our results show major changes in GitHub over the last decade:
    (1) the consolidation of \textit{README.md}, \textit{.gitignore}, and \textit{LICENSE} as standard artifacts;
    (2) the rise of GitHub Actions as the dominant CI/CD platform;
    (3) the growth of configuration formats such as TOML, YAML, and JSON, alongside a decline in XML;
    (4) new trends, such as the growth of \textit{Dockerfile}; and
    (5) emerging content related to LLMs and generative AI (e.g., AGENTS.md and CLAUDE.md).
    Based on our findings, we discuss implications, including that open source is not only evolving organically but also increasingly
    guided by GitHub’s standards, the rise and fall of technologies, and the potential support for mining software repository studies.
\end{abstract}


\section{Introduction}

GitHub is the largest code hosting platform, with millions of repositories spanning a wide range of technologies.
Repositories are commonly expected to follow software development best practices, including testing, CI/CD pipelines, documentation, license information, installation files, and contribution guidelines, to name a few~\cite{gh_healthy, gh_guidelines}.
Recently, repositories have also started incorporating generative LLMs and AI artifacts, such as configuration files for coding agents~\cite{fan2023large, hou2023large, agentminingpaper, robbes2026agentic}.

Despite this, little is known about the actual contents of GitHub's real-world repositories.
For example, although contribution guidelines are recommended~\cite{gh_healthy, gh_guidelines, elazhary2019not, falcucci2025contribution}, the extent to which they are actually adopted in practice remains unclear.
The same observation applies to other important artifacts of software development, such as documentation, CI/CD, and licensing.
Rather than assuming which software development artifacts are adopted, \textbf{practitioners, researchers, and educators should rely on empirical evidence of what developers actually include in their repositories}.
For example, practitioners could use such evidence to benchmark their projects against common practices, researchers could design more representative empirical studies, and educators could align their teaching with what developers actually do.

\begin{table}[t]
    \centering
    \caption{Overview of the dataset~\cite{gh_dataset}.}
    \begin{tabular}{lrrrr}
        \toprule
        \textbf{Year} & \textbf{Repositories} & \textbf{Files} & \textbf{Directories} & \textbf{Extensions} \\ \midrule
        2026          & 10,000                & 10,904,237     & 1,970,740            & 14,770              \\
        2021          & 6,371                 & 5,709674       & 1,025,852            & 10,901              \\
        2016          & 2,591                 & 2,650,192      & 402,451              & 9,487               \\
        \bottomrule
    \end{tabular}
    \label{tab:dataset}
\end{table}

This paper presents an initial empirical analysis to better understand the contents of real-world GitHub repositories.
Specifically, we analyze the files, directories, and extensions present in 10,000 GitHub repositories, as well as their evolution over ten years.
Table~\ref{tab:dataset} summarizes our dataset~\cite{gh_dataset}.
We propose two research questions to investigate the current state and the evolution of the repositories:

\begin{itemize}

    \item \textbf{RQ1. What files, directories, and extensions are commonly found in GitHub repositories?}
          Overall, we detected that the most common files in GitHub repositories are \textit{README.md}, \textit{.gitignore}, and \textit{LICENSE}.
          Commonly used Markdown files are related to open-source best practices, such as \textit{CONTRIBUTING.md}, \textit{CODE\_OF\_CONDUCT.md}, and \textit{SECURITY.md}.
          Also, most emerging content is related to LLMs and generative AI (e.g., \textit{CLAUDE.md} and \textit{AGENTS.md}).

    \item \textbf{RQ2. How do files, directories, and extensions in GitHub repositories evolve?}
          Over the last decade, we observed major changes in GitHub:
          (1) the consolidation of \textit{README.md}, \textit{.gitignore}, and \textit{LICENSE} as standard artifacts; (2) the rise of GitHub Actions as the dominant CI/CD platform; (3) the growth of configuration formats such as TOML, YAML, and JSON, alongside a decline in XML; (4) growth in Python and JavaScript extensions and a decline in C/C++, Java, and PHP; and (5) new trends, such as the growth of \textit{package.json} and \textit{Dockerfile}.

\end{itemize}

Based on our findings, we discuss implications for practitioners, researchers, and educators, including that open source is not only evolving organically but also increasingly guided by GitHub’s standards, the rise and fall of technologies, and the potential support for mining software repository studies.

\smallskip

\noindent\textbf{Contributions:}
The contributions of this study are threefold:
(1) we conduct a large-scale empirical study to explore the actual content of GitHub repositories;
(2) we derive actionable implications for practitioners, researchers, and educators; and
(3) we provide a dataset of files, directories, and extensions from 10K repositories to support further research~\cite{gh_dataset}.

\begin{table*}[t]
    \scriptsize
    \centering
    \caption{Files, directories, and extensions in 2026.}
    \begin{minipage}{0.32\linewidth}
        \centering
        \subcaption{Files}
        \begin{tabular}{rlrr}
            \toprule
            \textbf{Pos} & \textbf{File}   & \textbf{Repos} & \textbf{\%} \\
            \midrule
            1            & README.md       & 9,532          & 95.3        \\
            2            & .gitignore      & 9,498          & 95.0        \\
            3            & LICENSE         & 7,309          & 73.1        \\
            4            & package.json    & 3,333          & 33.3        \\
            5            & CONTRIBUTING.md & 3,170          & 31.7        \\
            6            & CHANGELOG.md    & 2,787          & 27.9        \\
            7            & .gitattributes  & 2,437          & 24.4        \\
            8            & index.html      & 2,392          & 23.9        \\
            9            & Dockerfile      & 2,390          & 23.9        \\
            10           & Makefile        & 2,341          & 23.4        \\
            \bottomrule
        \end{tabular}
    \end{minipage}\hfill
    \begin{minipage}{0.32\linewidth}
        \centering
        \subcaption{Directories}
        \begin{tabular}{rlrr}
            \toprule
            \textbf{Pos} & \textbf{Directory} & \textbf{Repos} & \textbf{\%} \\
            \midrule
            1            & .github            & 8,245          & 82.5        \\
            2            & workflows          & 7,730          & 77.3        \\
            3            & src                & 6,009          & 60.1        \\
            4            & docs               & 3,740          & 37.4        \\
            5            & tests              & 3,545          & 35.5        \\
            6            & scripts            & 3,060          & 30.6        \\
            7            & test               & 3,004          & 30.0        \\
            8            & ISSUE\_TEMPLATE    & 2,817          & 28.2        \\
            9            & utils              & 2,608          & 26.1        \\
            10           & assets             & 2,598          & 26.0        \\
            \bottomrule
        \end{tabular}
    \end{minipage}\hfill
    \begin{minipage}{0.32\linewidth}
        \centering
        \subcaption{Extensions}
        \begin{tabular}{rlrr}
            \toprule
            \textbf{Pos} & \textbf{Extension} & \textbf{Repos} & \textbf{\%} \\
            \midrule
            1            & md                 & 9,860          & 98.6        \\
            2            & gitignore          & 9,498          & 95.0        \\
            3            & yml                & 8,414          & 84.1        \\
            4            & json               & 6,920          & 69.2        \\
            5            & png                & 6,262          & 62.6        \\
            6            & txt                & 5,689          & 56.9        \\
            7            & sh                 & 4,681          & 46.8        \\
            8            & yaml               & 4,207          & 42.1        \\
            9            & js                 & 4,047          & 40.5        \\
            10           & py                 & 3,942          & 39.4        \\
            \bottomrule
        \end{tabular}
    \end{minipage}
    \label{tab:rq1-all}
\end{table*}

\section{Study Design}

\subsection{Selecting Repositories}

Our goal is to analyze real-world, actively maintained repositories hosted on GitHub.
To this end, we rely on the SEART GitHub Search Engine (seart-ghs), a tool that allows researchers to sample repositories to use for empirical studies by using multiple combinations of selection criteria~\cite{Dabic:msr2021data}.
This tool maintains metadata for all GitHub repositories with at least ten stars.
Based on seart-ghs, we selected the repositories that meet the following criteria: at least 100 commits, not being forks, having at least one commit in 2026, and having at least 100 stars (the star metric is primarily adopted in the software mining literature as a proxy of popularity~\cite{icsme2016, jss-2018-github-stars}).


This process yielded an initial set of 116,013 repositories, from which we randomly selected 10,000 for analysis. Random sampling was used to ensure a diverse set of repositories, rather than focusing only on the most popular ones, as would be the case if we had selected the top 10K.
On the median, the selected repositories have 211 stars and 557 commits.
They span 44 primary programming languages and include projects from organizations such as Microsoft, Google, and Facebook.

\subsection{Extracting Files and Directories}

Next, we relied on the \emph{GitHub REST APIs for Git trees}\footnote{\url{http://docs.github.com/en/rest/git/trees}} to collect all files, directories, and extensions from the selected repositories.
We collected data for the last 10 years, considering the snapshots from 2016, 2021, and 2026.
Table~\ref{tab:dataset} summarizes our dataset.
For 2026, we analyzed the 10,000 repositories comprising over 10.9 million files, 1.9 million directories, and 14,770 distinct extensions.
Next, we analyzed the repositories from the 2026 dataset that existed in 2021 and 2016, totalling 6,371 and 2,591 repositories, respectively.
Our dataset is publicly available~\cite{gh_dataset}.

\subsection{Research Questions}

We propose two research questions to explore the current state (RQ1) and the evolution of the repositories (RQ2).
The rationale is to identify which software development artifacts are adopted in practice and how the usage changes over time to support practitioners, researchers, and educators.
Practitioners can benchmark against common practices (e.g., typical repository structures), researchers can design more representative studies (e.g., building datasets that reflect current development artifacts), and educators can better align teaching with real-world development (e.g., motivating the usage of certain tools).
In addition, this study is a first step toward a better understanding of GitHub content at an ultra-large scale.

\section{Current Content of GitHub (RQ1)}

\subsection{Files, directories, and extensions}

Table~\ref{tab:rq1-all} provides an overview of the most commonly used files, directories, and extensions in 2026.
It is interesting to note that, in all cases, the top three entries appear with considerably high frequency.
Among the files, the three most frequent are \textit{README.md} (documentation), \textit{.gitignore} (version control), and \textit{LICENSE} (licensing).
Considering the directories, the three most common are \textit{.github} (repository configuration), \textit{workflows} (CI/CD), and \textit{src} (source code).
Lastly, regarding the extensions, the top three are: \textit{.md}, \textit{.gitignore}, and \textit{.yml}.

\begin{boxH}
    \textbf{Finding 1}:
    Currently, the most common files in GitHub repositories are \textit{README.md}, \textit{.gitignore}, and \textit{LICENSE}. The most common directory is \textit{.github}, and the most common extension is \textit{.md}.
\end{boxH}

\begin{table*}[t]
    \scriptsize
    \centering
    \caption{Emerging files, directories, and extensions in 2026.}
    \begin{minipage}{0.32\linewidth}
        \centering
        \subcaption{Emerging files}
        \begin{tabular}{rlrr}
            \toprule
            \textbf{Pos} & \textbf{File}           & \textbf{Repos} & \textbf{\%} \\
            \midrule
            1            & CLAUDE.md               & 898            & 9.0         \\
            2            & AGENTS.md               & 846            & 8.5         \\
            3            & SKILL.md                & 553            & 5.5         \\
            4            & eslint.config.mjs       & 505            & 5.0         \\
            5            & eslint.config.js        & 487            & 4.9         \\
            6            & uv.lock                 & 465            & 4.7         \\
            7            & vitest.config.ts        & 413            & 4.1         \\
            8            & tsconfig.node.json      & 381            & 3.8         \\
            9            & vite-env.d.ts           & 377            & 3.8         \\
            10           & copilot-instructions.md & 313            & 3.1         \\
            \bottomrule
        \end{tabular}
    \end{minipage}\hfill
    \begin{minipage}{0.32\linewidth}
        \centering
        \subcaption{Emerging directories}
        \begin{tabular}{rlrr}
            \toprule
            \textbf{Pos} & \textbf{Directory} & \textbf{Repos} & \textbf{\%} \\
            \midrule
            1            & skills             & 567            & 5.7         \\
            2            & .claude            & 450            & 4.5         \\
            3            & mcp                & 243            & 2.4         \\
            4            & .cursor            & 161            & 1.6         \\
            5            & .vitepress         & 122            & 1.2         \\
            6            & .agents            & 120            & 1.2         \\
            7            & llm                & 109            & 1.1         \\
            8            & [id]               & 83             & 0.8         \\
            9            & openai             & 82             & 0.8         \\
            10           & src-tauri          & 79             & 0.8         \\
            \bottomrule
        \end{tabular}
    \end{minipage}\hfill
    \begin{minipage}{0.32\linewidth}
        \centering
        \subcaption{Emerging extensions}
        \begin{tabular}{rlrr}
            \toprule
            \textbf{Pos} & \textbf{Extension} & \textbf{Repos} & \textbf{\%} \\
            \midrule
            1            & mts                & 224            & 2.2         \\
            2            & mdc                & 131            & 1.3         \\
            3            & slnx               & 121            & 1.2         \\
            4            & xcprivacy          & 108            & 1.1         \\
            5            & astro              & 82             & 0.8         \\
            6            & prisma             & 64             & 0.6         \\
            7            & codespellrc        & 61             & 0.6         \\
            8            & clangd             & 53             & 0.5         \\
            9            & work               & 48             & 0.5         \\
            10           & lycheeignore       & 46             & 0.5         \\
            \bottomrule
        \end{tabular}
    \end{minipage}
    \label{tab:rq1-novel}
\end{table*}

\subsection{Emerging files, directories, and extensions}

Table~\ref{tab:rq1-novel} presents the novel files, directories, and extensions identified in 2026.
Most of the emerging content in GitHub repositories is related to LLMs, generative AI, and coding agents.
In this context, we identified \textit{CLAUDE.md},\footnote{\url{https://code.claude.com/docs/en/best-practices}} \textit{AGENTS.md},\footnote{\url{https://agents.md}} \textit{SKILL.md},\footnote{\url{https://agentskills.io}} and \textit{copilot-instructions.md}\footnote{\url{https://docs.github.com/en/copilot/how-tos/copilot-on-github/customize-copilot/add-custom-instructions/add-repository-instructions}} for guiding coding agents.
For directories, we detected \textit{skills}, \textit{.claude}, \textit{mcp}, \textit{.cursor}, \textit{.agents}, and \textit{openai}.

It is also worth noting other emerging contents, such as \textit{eslint.config} (configuration file used by ESLint to define linting rules and project settings), \textit{uv.lock} (lockfile generated by uv that records the exact versions of Python dependencies), and \textit{vitest.config.ts} (configuration file used by Vitest to define testing settings). Regarding the extensions, the most frequent emerging extension is \textit{.mts}, a TypeScript file extension used for ECMAScript modules.

\begin{boxH}
    \textbf{Finding 2}:
    Most of the emerging files and directories in GitHub repositories are related to LLMs, generative AI, and coding agents such as \textit{CLAUDE.md}, \textit{AGENTS.md}, \textit{SKILL.md}, \textit{skills}, and \textit{.claude}.
\end{boxH}

\subsection{Markdown files}

The Markdown extension is the most common in GitHub repositories; therefore, we provide a detailed analysis of Markdown files in Table~\ref{tab:rq1-md}.
Most of the top Markdown files are related to open-source best practices, including \textit{README.md}, \textit{CONTRIBUTING.md}, \textit{CHANGELOG.md}, \textit{CODE\_OF\_CONDUCT.md}, \textit{SECURITY.md}, \textit{LICENSE.md}, and \textit{PULL\_REQUEST\_TEMPLATE.md}.

GitHub itself recommends the adoption of such files so that repository maintainers can establish guidelines that help collaborators make meaningful contributions~\cite{gh_healthy}.
For example, GitHub recommends creating contributing guidelines~\cite{gh_guidelines}, code of conduct~\cite{gh_code_of_conduct}, and license~\cite{gh_license} files.

\begin{table}[h]
    \centering
    \caption{Markdown (\textit{.md}) files in 2026.}
    \begin{tabular}{rlrr}
        \toprule
        \textbf{Pos} & \textbf{File}        & \textbf{Repos} & \textbf{\%} \\
        \midrule
        1            & README.md            & 9,532          & 95.3        \\
        2            & CONTRIBUTING.md      & 3,170          & 31.7        \\
        3            & CHANGELOG.md         & 2,787          & 27.9        \\
        4            & CODE\_OF\_CONDUCT.md & 1,625          & 16.2        \\
        5            & SECURITY.md          & 1,354          & 13.5        \\
        \bottomrule
    \end{tabular}
    \label{tab:rq1-md}
\end{table}

\begin{boxH}
    \textbf{Finding 3}:
    The most commonly used Markdown files are related to open-source best practices, including \textit{CONTRIBUTING.md}, \textit{CHANGELOG.md}, \textit{CODE\_OF\_CONDUCT.md}, and \textit{SECURITY.md}.
\end{boxH}

\subsection{Dotfiles}

Dotfiles are hidden configuration files whose names start with a dot.
Table~\ref{tab:rq1-dotfiles} presents the most common dotfiles found in GitHub repositories.
The majority of the dotfiles are Git-related, including \textit{.git\-ignore}, \textit{.git\-attributes}, \textit{.git\-modules}, \textit{.pre-\-commit-\-config.\-yaml}, and \textit{.gitkeep}.
We also identified ignore-related files, such as \textit{.git\-ignore}, \textit{.docker\-ignore}, and \textit{.prettier\-ignore}, which are used to exclude files and directories from Git tracking, Docker build contexts, and Prettier code formatting.

\begin{table}[h]
    \centering
    \caption{Dotfiles in 2026.}
    \begin{tabular}{rlrr}
        \toprule
        \textbf{Pos} & \textbf{File}  & \textbf{Repos} & \textbf{\%} \\
        \midrule
        1            & .gitignore     & 9,498          & 95.0        \\
        2            & .gitattributes & 2,437          & 24.4        \\
        3            & .editorconfig  & 2,013          & 20.1        \\
        4            & .dockerignore  & 1,283          & 12.8        \\
        5            & .gitmodules    & 995            & 9.9         \\
        \bottomrule
    \end{tabular}
    \label{tab:rq1-dotfiles}
\end{table}

\begin{boxH}
    \textbf{Finding 4}:
    The majority of the dotfiles are Git-related, such as \textit{.gitignore}, \textit{.gitattributes}, and \textit{.gitmodules}.
    \end{boxH}

\subsection{Files without extension}

Files without an explicit extension are also commonly found in GitHub repositories, as detailed in Table~\ref{tab:rq1-no-ext}.
The most common are: \textit{LICENSE}, \textit{Dockerfile}, and \textit{Makefile}.
Here, we observe two main groups of files: those related to build and automation processes (e.g., \textit{Dockerfile}, \textit{Makefile}, and \textit{gradlew}) and those related to documentation and project governance (e.g., \textit{LICENSE}, \textit{CODEOWNERS}, and \textit{AUTHORS}).

\begin{table}[h]
    \centering
    \caption{Files without extension in 2026.}
    \begin{tabular}{rlrr}
        \toprule
        \textbf{Pos} & \textbf{File} & \textbf{Repos} & \textbf{\%} \\
        \midrule
        1            & LICENSE,      & 7,309          & 73.1        \\
        2            & Dockerfile    & 2,390          & 23.9        \\
        3            & Makefile      & 2,341          & 23.4        \\
        4            & CODEOWNERS    & 1,017          & 10.2        \\
        5            & gradlew       & 700            & 7.0         \\
        \bottomrule
    \end{tabular}
    \label{tab:rq1-no-ext}
\end{table}

\begin{boxH}
    \textbf{Finding 5}:
    The most common files without explicit extensions in GitHub repositories are related to build and automation processes (e.g., \textit{Makefile}) or documentation and project governance (e.g., \textit{LICENSE}).
\end{boxH}

\newpage
\section{Content Evolution of GitHub (RQ2)}

\subsection{Overview}

In this research question, we investigate the evolution of the content in GitHub repositories.
Table~\ref{tab:summary} summarizes the median number of files, directories, and file extensions per repository.
Overall, we observe that repositories tend to grow over time, increasing both the number of files (from 111 in 2016 to 211 in 2026) and directories (from 20 to 42).
Interestingly, the median number of file extensions per repository also increased, from 12 in 2016 to 17 in 2026, suggesting greater diversity in the technologies being used.


\begin{table}[h]
    \centering
    \caption{Overview of the content evolution (median).}
    \begin{tabular}{lrrrr}
        \toprule
        \textbf{Year} & \textbf{Repos.} & \textbf{Files} & \textbf{Directories} & \textbf{Extensions} \\
        \midrule
        2026          & 10,000          & 211            & 42                   & 17                  \\
        2021          & 6,371           & 131            & 27                   & 14                  \\
        2016          & 2,591           & 111            & 20                   & 12                  \\
        \bottomrule
    \end{tabular}
    \label{tab:summary}
\end{table}

\begin{boxH}
    \textbf{Finding 6}:
    Overall, repositories have grown over the last decade in terms of the number of files, directories, and diversity of file extensions.
\end{boxH}

\subsection{Increasing/decreasing files, directories, and extensions}

Table~\ref{tab:rq2a} presents the most common files with increasing prevalence, while Table~\ref{tab:rq2b} shows the most common files with decreasing prevalence from 2016 to 2026.
For example, the prevalence of \textit{README.md} files increased from 77.8\% of repositories in 2016 to 95.3\% in 2026 (+17.5\%).
Other files with a large increase include \textit{.gitignore} (from 84.5\% to 95.0\%, +10.5\%), \textit{LICENSE} (from 50.8\% to 73.1\%, +22.3\%), \textit{package.json} (from 17.2\% to 33.3\%, +16.1\%), and \textit{Dockerfile} (from 5.7\% to 23.9\%, +18.2\%).
Regarding the decreasing files, the most notable decline occurs for \textit{.travis.yml}, which decreased from 46.4\% in 2016 to 5.3\% in 2026 (-41.1\%).
We recall that Travis CI was the dominant CI/CD platform before the rise of GitHub Actions~\cite{decan2022use}.



Tables~\ref{tab:rq2c} and \ref{tab:rq2d} list the most popular directories with increasing and decreasing prevalence, respectively.
Two directories stand out with a remarkable increase in the last 10 years: \textit{.github} (from 5.3\% to 82.5\%, +77.2\%) and \textit{workflows} (from 0.2\% to 77.3\%, +77.1\%).
Both are responsible for setting up CI/CD pipelines on GitHub, indicating a significant increase in this practice in the last decade~\cite{decan2022use}.




Tables~\ref{tab:rq2e} and \ref{tab:rq2f} feature the most common extensions with increasing and decreasing prevalence, respectively.
Regarding the increasing extensions, \textit{.yaml} (+35.9\%), \textit{.toml} (+33.0\%), \textit{.yml} (+30.9\%), and \textit{.json} (+30.3\%) presented the highest increases.
All these extensions are commonly used to configure the development environment, including CI/CD pipelines, linters, and package managers~\cite{alfadel2023empirical, Montandon2019, decan2022use}.
In contrast, the \textit{.xml} extension decreased in usage by 6.4\%.
Interestingly, file extensions with decreasing prevalence are centered on source code files of traditional programming languages.
For example, \textit{.c}, \textit{.h}, and \textit{.in} extensions are the top-3 most affected ones, with 8.3\%, 7.2\%, and 7.0\% reductions; these files are typically used in C/C++ projects.
Source code extensions related to Java and PHP (i.e.,~\textit{.java} and \textit{.php}) decreased by 6.2\% and 4.4\%.
In contrast, files related to Python (\textit{.py}) and JavaScript (\textit{.js}) increased by 10\% and 7.5\%.


\begin{boxH}
    \textbf{Finding 7}:
    Over the last decade, we observed major changes in GitHub:
    \textbf{(1)} the consolidation of \textit{README.md}, \textit{.gitignore}, and \textit{LICENSE} as standard artifacts; \textbf{(2)} the rise of GitHub Actions as the dominant CI/CD platform (and the decline of Travis CI); \textbf{(3)} the growth of configuration formats such as TOML, YAML, and JSON, alongside a decline in XML; \textbf{(4)} growth in Python and JavaScript extensions and a decline in C/C++, Java, and PHP; and \textbf{(5)} new trends, such as the growth of \textit{package.json} and \textit{Dockerfile}.
\end{boxH}

\begin{table*}[t]
    \scriptsize
    \centering
    \caption{Increasing and decreasing files, directories, and extensions over time (2016--2026).}
    \begin{minipage}{0.32\linewidth}
        \centering
        \subcaption{Increasing files (\%)}
        \label{tab:rq2a}
        \begin{tabular}{rlrr|r}
            \toprule
            \textbf{Pos} & \textbf{File}   & \textbf{2016} & \textbf{2026} & $\Delta$ \\ \midrule          
            1            & README.md       & 77.8          & 95.3          & 17.5     \\         
            2            & .gitignore      & 84.5          & 95.0          & 10.5     \\         
            3            & LICENSE         & 50.8          & 73.1          & 22.3     \\         
            4            & package.json    & 17.2          & 33.3          & 16.1     \\         
            5            & CONTRIBUTING.md & 15.7          & 31.7          & 16.0     \\         
            6            & CHANGELOG.md    & 12.6          & 27.9          & 15.3     \\         
            7            & .gitattributes  & 14.4          & 24.4          & 10.0     \\         
            8            & index.html      & 18.4          & 23.9          & 5.5      \\         
            9            & Dockerfile      & 5.7           & 23.9          & 18.2     \\         
            10           & \_\_init\_\_.py & 15.9          & 22.3          & 6.4      \\         
            \bottomrule
        \end{tabular}
    \end{minipage}\hfill\hfill
    \vspace{1em}
    \begin{minipage}{0.32\linewidth}
        \centering
        \subcaption{Increasing directories (\%)}
        \label{tab:rq2c}
        \begin{tabular}{rlrr|r}
            \toprule
            \textbf{Pos} & \textbf{Directory} & \textbf{2016} & \textbf{2026} & $\Delta$ \\ \midrule 
            1            & .github            & 5.3           & 82.5          & 77.2     \\   
            2            & workflows          & 0.2           & 77.3          & 77.1     \\  
            3            & src                & 46.7          & 60.1          & 13.4     \\  
            4            & docs               & 15.4          & 37.4          & 22.0     \\  
            5            & tests              & 25.5          & 35.5          & 10.0     \\  
            6            & scripts            & 14.2          & 30.6          & 16.4     \\  
            7            & ISSUE\_TEMPLATE    & 0.1           & 28.2          & 28.1     \\  
            8            & utils              & 10.5          & 26.1          & 15.6     \\  
            9            & assets             & 9.1           & 26.0          & 16.9     \\   
            10           & images             & 16.3          & 23.9          & 7.6      \\ 
            \bottomrule
        \end{tabular}
    \end{minipage}\hfill
    \begin{minipage}{0.32\linewidth}
        \centering
        \subcaption{Increasing extensions (\%)}
        \label{tab:rq2e}
        \begin{tabular}{rlrr|r}
            \toprule
            \textbf{Pos} & \textbf{Extension} & \textbf{2016} & \textbf{2026} & $\Delta$ \\ \midrule  
            1            & md                 & 83.4          & 98.6          & 15.2     \\ 
            2            & gitignore          & 84.5          & 95.0          & 10.5     \\ 
            3            & yml                & 53.2          & 84.1          & 30.9     \\ 
            4            & json               & 38.9          & 69.2          & 30.3     \\ 
            5            & png                & 43.0          & 62.6          & 19.6     \\ 
            6            & sh                 & 38.0          & 46.8          & 8.8      \\ 
            7            & yaml               & 6.2           & 42.1          & 35.9     \\ 
            8            & js                 & 33.0          & 40.5          & 7.5      \\ 
            9            & py                 & 29.4          & 39.4          & 10.0     \\ 
            10           & toml               & 2.6           & 35.6          & 33.0     \\ 
            \bottomrule
        \end{tabular}
    \end{minipage}\hfill
    \begin{minipage}{0.32\linewidth}
        \centering
        \subcaption{Decreasing files (\%)}
        \label{tab:rq2b}
        \begin{tabular}{rlrr|r}
            \toprule
            \textbf{Pos} & \textbf{File} & \textbf{2016} & \textbf{2026} & $\Delta$ \\ \midrule 
            1            & .travis.yml   & 46.4          & 5.3           & -41.1    \\ 
            2            & Makefile      & 24.7          & 23.4          & -1.3     \\ 
            3            & README        & 19.5          & 7.0           & -12.5    \\ 
            4            & LICENSE.txt   & 14.7          & 10.6          & -4.1     \\ 
            5            & setup.py      & 12.7          & 7.6           & -5.1     \\ 
            6            & COPYING       & 12.7          & 4.4           & -8.3     \\ 
            7            & README.txt    & 10.8          & 4.8           & -6.0     \\ 
            8            & AUTHORS       & 10.5          & 3.9           & -6.6     \\ 
            9            & pom.xml       & 8.7           & 4.5           & -4.2     \\ 
            10           & appveyor.yml  & 7.6           & 1.8           & -5.8     \\ 
            \bottomrule
        \end{tabular}
    \end{minipage}\hfill
    \begin{minipage}{0.32\linewidth}
        \centering
        \subcaption{Decreasing directories (\%)}
        \label{tab:rq2d}
        \begin{tabular}{rlrr|r}
            \toprule
            \textbf{Pos} & \textbf{Directory} & \textbf{2016} & \textbf{2026} & $\Delta$ \\ \midrule 
            1            & test               & 34.0          & 30.0          & -4.0     \\  
            2            & doc                & 16.3          & 9.0           & -7.3     \\  
            3            & bin                & 14.4          & 12.4          & -2.0     \\  
            4            & java               & 13.4          & 11.3          & -2.1     \\  
            5            & css                & 13.0          & 12.0          & -1.0     \\  
            6            & util               & 12.9          & 11.8          & -1.1     \\  
            7            & js                 & 11.9          & 9.8           & -2.1     \\  
            8            & include            & 10.3          & 7.8           & -2.5     \\  
            9            & source             & 7.7           & 6.8           & -0.9     \\  
            10           & org                & 7.4           & 4.0           & -3.4     \\  
            \bottomrule
        \end{tabular}
    \end{minipage}\hfill
    \begin{minipage}{0.32\linewidth}
        \centering
        \subcaption{Decreasing extensions (\%)}
        \label{tab:rq2f}
        \begin{tabular}{rlrr|r}
            \toprule
            \textbf{Pos} & \textbf{Extension} & \textbf{2016} & \textbf{2026} & $\Delta$ \\ \midrule 
            1            & xml                & 33.1          & 26.7          & -6.4     \\ 
            2            & h                  & 25.4          & 18.2          & -7.2     \\ 
            3            & c                  & 20.9          & 12.6          & -8.3     \\ 
            4            & in                 & 19.5          & 12.5          & -7.0     \\ 
            5            & java               & 17.0          & 10.8          & -6.2     \\ 
            6            & cpp                & 15.2          & 12.2          & -3.0     \\ 
            7            & properties         & 14.2          & 13.1          & -1.1     \\ 
            8            & cfg                & 12.5          & 8.4           & -4.1     \\ 
            9            & rst                & 11.4          & 8.4           & -3.0     \\ 
            10           & php                & 10.2          & 5.8           & -4.4     \\ 
            \bottomrule
        \end{tabular}
    \end{minipage}
    \label{tab:rq2}
\end{table*}

\section{Discussion and Implications}

\subsection{Current state of GitHub}

Nowadays, a typical GitHub repository contains \textit{README.md} (growth from 77.8\% to 95.3\%; +17.5\%), \textit{.git\-ignore} (84.5\%--95\%; +10.5\%), and \textit{LICENSE} (50.8\%--73.1\%; +22.3\%) files.
Interestingly, these files are suggested by the GitHub platform when creating a new repository~\cite{new-repo} and can be automatically generated, which may help explain their prevalence in practice.
Other highly prevalent artifacts (each above 70\%) include the \textit{.github} and \textit{workflows} directories, indicating widespread adoption of GitHub Actions for CI/CD, even though GitHub Actions is not currently suggested by GitHub when creating a new repository.

\vspace*{0.2cm}
\noindent\textbf{\textsc{Implication \#1.}}
These findings highlight the strong influence of GitHub in shaping open source.
In practice, open source is not only evolving organically, but also increasingly guided by the platform's standards.
Given the widespread use of artifacts such as \textit{workflow} and \textit{ISSUE\_TEMPLATE} directories, GitHub could extend support for new repositories with CI/CD and issue management templates.

\subsection{Rise and fall of technologies}

Our study allows quantifying how the usage of certain artifacts evolves.
For example, we observed the decline of Travis CI and the rise of GitHub Actions~\cite{decan2022use}.
Similarly, we identified that lightweight data formats, such as TOML, YAML, and JSON, increased in favor of traditional ones, like XML.
There are also specific cases, such as the decline of \textit{setup.py} (-5.1\%) (build scripts for Python projects), possibly in favor of more modern alternatives as \emph{pyproject.toml} (+11.5\%).

\vspace*{0.2cm}
\noindent\textbf{\textsc{Implication \#2.}}
At first glance, these trends suggest that maintainers are continuously adapting their repositories to new technologies.
Researchers can further investigate the reasons for such changes, as well as propose solutions to support these transitions automatically~\cite{Macedo2025}.
This information can also be used by educators and practitioners to guide the technology selection for teaching and adoption.


\subsection{Support to mining software repository studies}

Many mining software repository studies rely on detecting specific files and directories.
For instance, to study coding agents, researchers must first identify repositories containing files such as \textit{AGENTS.md} and \textit{CLAUDE.md}~\cite{robbes2026agentic, agentminingpaper, lulla2026impact, Santos2026, hora2026coding, galster2026configuring, gloaguen2026evaluating}.
Similarly, studies on contribution guidelines depend on detecting repositories with \textit{CONTRIBUTING.md} files~\cite{elazhary2019not, falcucci2025contribution, tsay2014influence}.
There is a plethora of other files that can support software mining tasks, such as \textit{LICENSE} for licensing~\cite{vendome2017license}, \textit{CHANGELOG.md} for break changes~\cite{xavier2017historical, brito2018and, hora2015developers}, \textit{package.json} and \textit{pom.xml} for dependencies~\cite{decan2019package, pinckney2023large, deAlcantaraJunior2026, Montandon2019}, \textit{setup.py} and \emph{pyproject.toml} for Python packaging~\cite{alfadel2023empirical}, \textit{Dockerfile} for containerization~\cite{henkel2020dataset, cito2017empirical}, to name a few.

\vspace*{0.2cm}
\noindent\textbf{\textsc{Implication \#3.}}
This study is a first step toward a better understanding of GitHub content at a large scale and in continuously updated settings.
We plan to develop a taxonomy of repository artifacts, e.g., common files related to software testing, CI/CD, and software dependencies.
This taxonomy can guide researchers in selecting an initial set of artifacts to consider when mining software repositories.

\section{Threats to Validity}

We analyzed files, directories, and extensions from a random sample of 10,000 open-source GitHub repositories that are actively maintained in 2026 and have at least 100 stars.
Thus, our findings may not directly generalize to other contexts, such as less actively maintained projects, less popular projects, closed-source projects, or other code-hosting platforms like GitLab or Bitbucket.

\section{Related Work}

%
As a key host of open-source software development, GitHub remains a primary source of empirical studies in SE, powering more than 70\% of MSR studies~\cite{vidoni2022systematic}.
At a meta-level, many studies have analyzed GitHub project content to report on the challenges of conducting SE research using its data~\cite{Kalliamvakou2016} and to recommend data curation strategies~\cite{munaiah2017curating,vidoni2022systematic} and tooling~\cite{Dabic:msr2021data}.
In comparison, fewer studies have focused on characterizing repository metadata as we have.
Gonzalez et al. characterized a decade of AI and ML repositories on GitHub, surfacing unique structural and workflow patterns in this community~\cite{gonzalez2020mlstate}.
Further studies have used repository content to investigate specific software practices, focusing on project \textsc{readme} documentation~\cite{prana2019readme}, contribution guidelines~\cite{elazhary2019not}, licensing~\cite{wu2024license}, and dependency management.
In contrast to these analyses of specific repository content, our study takes a broad, cross-cutting view of \emph{all} files, directories, and extensions across 10{,}000 actively maintained repositories, and tracks their evolution over ten years.


\section{Conclusion and Further Steps}

This paper presented an initial empirical analysis to better understand the contents of real-world GitHub repositories.
We analyzed the files, directories, and extensions present in 10,000 GitHub repositories, as well as their evolution.
In short, our results revealed major changes in GitHub over the last decade.


\smallskip

\noindent\textbf{Further Steps:}
As future work, we plan to extend this analysis to an ultra-large scale, refine our analysis by better characterizing data across domains, and develop a web platform that enables practitioners, researchers, and educators to easily track the prevalence of files, directories, and extensions.

\section*{Acknowledgments}

This research was supported by CNPq (process 403304/2025-3), CAPES, and FAPEMIG.

\bibliographystyle{IEEEtran}
\bibliography{main}

\end{document}